# Photonic Cyclone: spatiotemporal optical vortex with controllable transverse orbital angular momentum

Andy Chong [1,2,3]*, Chenhao Wan [1,4], Jian Chen [1], and Qiwen Zhan [1,3]*

[1] School of Optical-Electrical and Computer Engineering, University of Shanghai for Science and Technology, 200093 Shanghai, China

[2] Department of Physics, University of Dayton, 300 College Park, Dayton, Ohio 45469, USA

[3] Department of Electro-Optics and Photonics, University of Dayton, 300 College Park, Dayton, Ohio 45469, USA

[4] School of Optical and Electronic Information, Huazhong University of Science and Technology, Wuhan, Hubei 430074, China

*Correspondence to: achong1@udayton.edu, qzhan1@udayton.edu

**Today, it is well known that light possesses a linear momentum which is along the propagation direction. Besides, scientists also discovered that light can possess an angular momentum (AM), a spin angular momentum (SAM) associated with circular polarization and an orbital angular momentum (OAM) owing to the azimuthally dependent phase. Even though such angular momenta are longitudinal in general, a SAM transverse to the propagation has opened up a variety of key applications [1]. In contrast, investigations of the transverse OAM are quite rare due to its complex nature. Here we demonstrate a simple method to generate a three dimensional (3D) optical wave packet with a controllable purely transverse OAM. Such a wave packet is a spatiotemporal (ST) vortex, which resembles an advancing cyclone, with optical energy flowing in the spatial and temporal dimension. Contrary to the transverse SAM, the magnitude of the transverse OAM carried by the photonic cyclone is scalable to a larger value by simple adjustments. Since the ST vortex carries a controllable OAM in the unique transverse dimension, it has a strong potential for novel applications that may not be possible otherwise. The scheme reported here can be readily adapted for the other spectra regime and different wave fields, opening tremendous opportunities for the study and applications of ST vortex in much broader scopes.**

Since there are only two SAM states of right and left circular polarization, the optical SAM is fundamentally limited to two values of $\pm\hbar$ per photon [2]. It is also well verified that the direction of such a SAM is longitudinal to the light propagation. In contrast, the transverse SAM occurs only in some special cases such as tightly focused beams [3] and evanescent waves of tightly confined waveguides [4]. Even though the transverse SAM is still limited to $\pm\hbar$ per photon, its unique direction can be utilized in emerging applications such as spin-orbit angular momentum coupling [5-7], quantum optical telecommunications [8], optical tweezing [9,10], etc.

On the other hand, the optical OAM behaves quite differently. The best known optical OAM occurs by applying a spirally increasing (decreasing) phase of $\exp(il\phi)$ on the transverse plane where $\phi$ is the azimuthal angle and $l$ is an integer referred to as a topological charge [11]. With such a spiral phase, the optical beam converts into a ring shape (Fig. 1a) with the transverse component of Poynting vectors, and therefore the transverse component of the energy circling around the phase singularity at the center [12]. Such an optical beam is referred to as an optical vortex since it resembles vortices in the fluid such as whirlpools, tornadoes, etc. The optical vortex exhibits a longitudinal adjustable OAM of $\pm l\hbar$ per photon [11]. Since the amount OAM can be many times larger than that of a SAM by tuning the topological charge, applications of the optical OAM are prolific in optical tweezers [13], super-resolution microscopy [14], optical communications [15], etc.

Even though the research of the longitudinal OAM is growing rapidly, the investigation of the transverse OAM has been lagged significantly. Limited numbers of theoretical works pointed out that a transverse OAM is possible in forms of ST vortices [16, 17]. In a nonlinear interaction of an extremely high power pulse and air, it was demonstrated that a small portion of the energy circulates in the meridional plane to form a ST vortex with a transverse OAM [18]. Despite the experimental success, a complicated nonlinear interaction is necessary. Furthermore, it is not suitable for practical applications since the leftover energy with zero OAM is dominant to wash out the small transverse OAM effect.

The limited outcome of the transverse OAM research is due to its complexity which the ST vortex inevitably has. To illustrate the point, let's examine the simplest ST vortex one can imagine. A typical 3D intensity profile of a pulsed vortex beam with the longitudinal OAM as shown in Fig. 1b. A spatial vortex has a spiral phase in the $x - y$ plane with a phase singularity of zero intensity at the center. If it is rotated by 90° with respect to a spatial axis ($x$ - axis in the figure), a 3D wave packet with a transverse OAM will be formed as shown in Fig. 1c. The resulting wave packet has a spiral phase in the meridional plane ($x - t$ plane). Such a wave packet is a ST vortex since the longitudinal component of Poynting vectors and therefore the optical energy circulates spatially and temporally. The wave packet, which travels with a vortex structure in a meridional plane, resembles an advancing wind vortex such as a cyclone. In this photonic cyclone, the spatial and temporal profiles are strongly coupled: the pulse profile varies significantly according to the spatial location. Unfortunately, there is no reliable method to generate such complicated spatiotemporally coupled optical wave packets up to date.

Here we demonstrate a photonic cyclone as a ST vortex by applying a spiral phase in the spatial frequency - frequency domain. It is nonintuitive to generate a ST vortex because it seems impossible to apply a spiral phase in the $x - t$ plane of a pulse. Since the angular momentum is a conserved physical property, it is intriguing to know whether a spiral phase will be conserved after two-dimensional (2D) Fourier transform from the spatial frequency-frequency domain ($k_x - \omega$ plane) to the spatial-temporal domain ($x - t$ plane). Suppose an optical field in the $k_x - \omega$ domain is given by $g_R(r)$ where $(r, \theta)$ are the corresponding polar coordinates. After a spiral phase of $e^{il\theta}$ is applied, a 2D Fourier transformed gives the field in $x - t$ domain [19]:

$$G(\rho, \phi) = F.T.\{g_R(r)e^{il\theta}\} = 2\pi\,(-i)^l e^{il\phi} H_l\{g_R(r)\},$$

where $(\rho, \phi)$ are Fourier conjugate polar coordinates, $H_l\{g_R(r)\} = \int_0^\infty r\ g_R(r)\, J_l(2\pi\rho r) dr$ and $J_l$ is the Bessel function of the first kind. It is crucial to notice that the spiral phase and its related optical OAM are indeed conserved after a Fourier transform. The theory indicates that a ST vortex can also be created by applying a spiral phase in the $k_x - \omega$ plane instead of the $x - t$ plane.

To apply a spiral phase in $k_x - \omega$ domain, a pulse shaper [20] with a 2D SLM with a phase-only control similar to Ref 21 and 22 is used (Fig. S2). Starting from chirped mode-locked pulses of ~3 ps durations, a diffraction grating and a cylindrical lens disperse frequencies spatially which act as a time to frequency Fourier transform. A spiral phase on the SLM (Fig. S2b) and an inverse Fourier transform by recollecting dispersed frequencies with a grating-cylindrical lens pair form a chirped ST vortex. A bit deviated from the theory, even though time-frequency Fourier transforms are executed by a grating-lens pair, a spatial Fourier transform procedure is not in present. However, a short free-space propagation within the pulse shaper gives sufficient diffraction, which acts as a sufficient spatial Fourier transform effect, to from ST vortices.

The phase measurement method is schematically shown in Fig. S3. By overlapping the chirped ST vortex with a short reference pulse (~90 fs) at an angle, interference fringes are formed. In fact, the ST vortex is intentionally chirped to be much longer than the reference pulse so that the fringe pattern represents the spatial phase profile of the thin temporal slice where the reference pulse overlaps. As the reference pulse is scanned in time, the full ST phase of the vortex can be reconstructed by collecting phase slices. For the ST vortex, a 3D diagnostic is essential to examine its spatiotemporally coupled nature. We used a simple 3D measurement technique which can be conveniently implemented in the experimental setup [23, 24].

Fig. 2 presents the phase measurement of the ST vortex with $l = 1$. The angle between the vortex and the reference beam is adjusted to form vertical fringes. The evolution of the fringe pattern can be understood as follows. As the reference pulse is overlapped at the head of the vortex, since the phase is almost constant, smooth continuous vertical fringes are observed. As the reference pulse scans toward the center of the vortex, the phase difference between the upper and lower part increases. Consequently, the upper fringes and the lower fringes are shifted with the center fringes start to bend to connect them. At the center of the vortex with $l = 1$, the up and down phase difference is π. Therefore, the up and down fringes are shifted by π while the center fringes diminish due to the zero intensity singularity. Just as the reference pulse passes the center, the phase difference becomes larger than π. The up and down fringes are still shifted but the center fringes bend in the opposite direction to connect them. As the reference pulse advances more toward the tail of the wave packet, fringes become vertical again. The experimentally observed phase behavior agrees very well with numerical simulations (Fig. 2b,c). Video S1 and S2 provide vivid visual representations of theoretical and experimental fringe patterns at various temporal locations for $l = 1$. Its ST phase is also measured and presented in Fig.2d.

The transverse OAM can be conveniently controlled by adjusting the topological charge on the SLM. For example, the transverse OAM direction can be reversed by simply applying $l = -1$ (Video S3). The amount of the OAM can be increased by applying larger topological charges such $l = 2$ (Fig.3). The $l = 2$ ST vortex exhibits interesting behavior. Since the up and down phase difference at the center for $l = 2$ is 2π (Fig. 3a, Fig. S4), the wave packet is merged at the center even under a very small propagation within the pulse shaper. Under such a circumstance, the $l = 2$ ST vortex will evolve into two $l = 1$ vortices quickly. Interestingly enough, this phenomena is also well observed in vortex beams as the instability of higher-order vortices with respect to perturbations [25]. Such behavior is well presented in the measurement which agrees very well with the theoretical prediction (Fig.3b,c, Video S4, S5).

Theoretical and experimental 3D intensity profiles are shown in Fig. 4. Video S6 and S7 show the experimental 3D profiles of the $l = 1$ and $l = 2$ ST vortex in every direction. While the theory and the experiment agree in general, the measurements show that the pulse head is slightly stretched due to asymmetric spectrum of the pulse source. For $l = 1$, a ST hole owing to the phase singularity clearly appears. Agreeing with the theory again, for $l = 2$, two ST holes appears in its 3D intensity profile. It is noteworthy that even though the $l = 2$ vortex decays to two $l = 1$ vortices, the total topological charge possessed by the wave packet is 2. Even though vortices are separated, it will effectively work as a single vortex with $l = 2$ for materials with response times longer than the separation between vortices (~600 fs). Currently, the length of the ST vortex is in the picosecond scale since it is temporally stretched. As the $l = 2$ ST vortex is dechirped to a much shorter pulse duration (~ 100 fs scale), the separation between vortices can be reduced significantly.

In conclusion, we demonstrated a photonic cyclone as a ST vortex with transverse controllable OAM. Even though the ST vortex has a spatiotemporally coupled profile, it was conveniently achieved by applying a spiral phase in a pulse shaper. Since the photonic cyclone has tunable OAM in the unique transverse direction, it is strongly believed that it will find special applications such as spin-orbit angular momentum coupling, quantum optics, etc. Finally, we want to note that the scheme reported here can be readily adapted for the other spectra regime as well as many other fields that involve wave phenomena (such as acoustics, electron beam, X-ray and so on), opening tremendous opportunities for the study and applications of ST vortex in much broader scopes.

**Author contribution**

A.C. proposed the original idea and performed all experiments and some theoretical analysis. C. W. performed all theoretical analysis and some experiments. A.C. and C.W. equally contributed to the project as first authors. J.C. contributed in developing the measurement method. Q. Z. guided the theoretical analysis and supervised the project. All authors contributed to writing the manuscript.

**Competing financial interest**

The authors declare no competing financial interests.

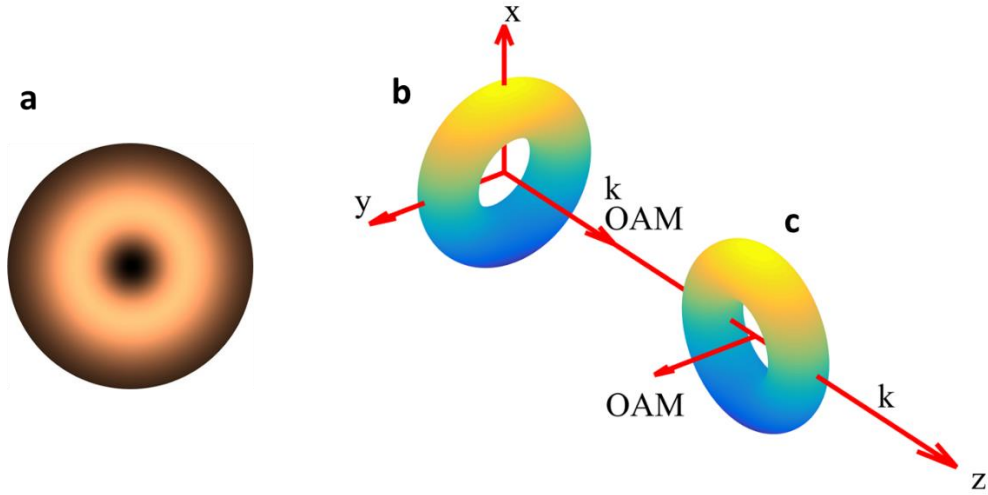

Figure 1. a) Typical vortex beam with a spiral phase of $\exp(il\theta)$ in the $x - y$ plane where $\theta = \tan^{-1}\left(\frac{y}{x}\right)$ with $l = 1$, b) A typical 3D profile of a pulse vortex beam with the longitudinal OAM, c) 3D profile of a ST vortex, which has the transverse OAM, by rotating the pulsed vortex beam in b.

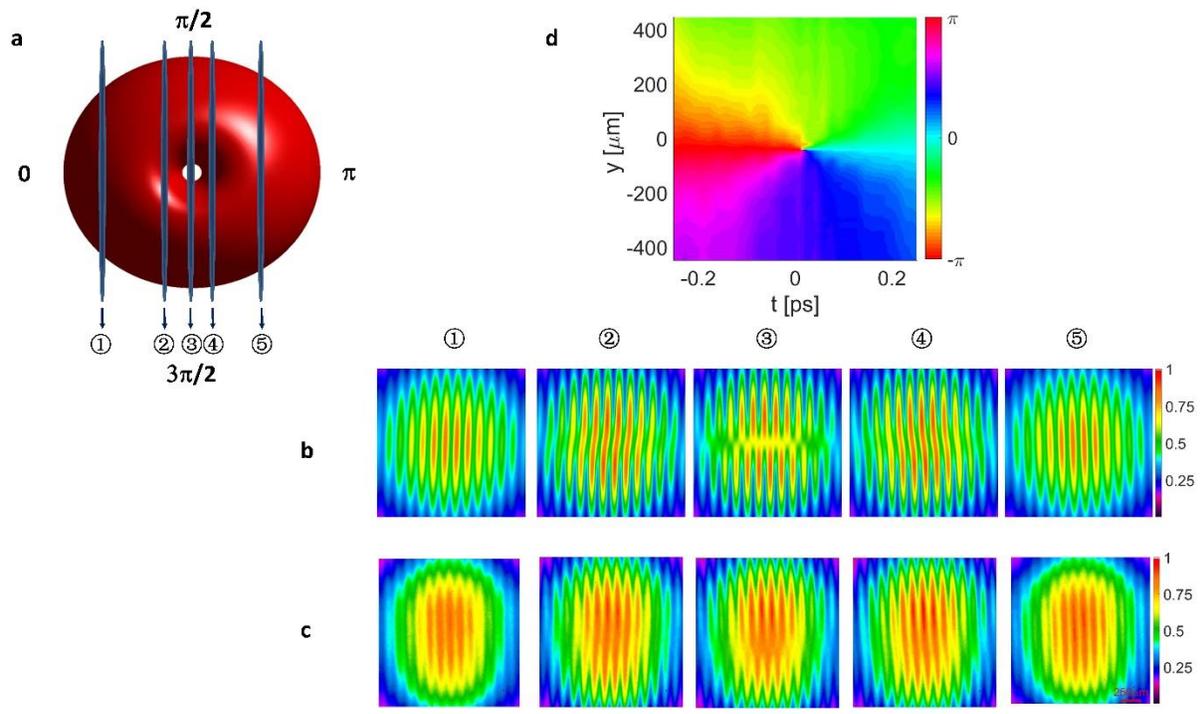

Figure 2. Measurement of the ST vortex for $l = 1$, a) Theoretical ST vortex profile, b) Theoretical phase fringe patterns at various temporal locations, c) Experimental fringe patterns at various locations, d) Reconstructed ST phase.

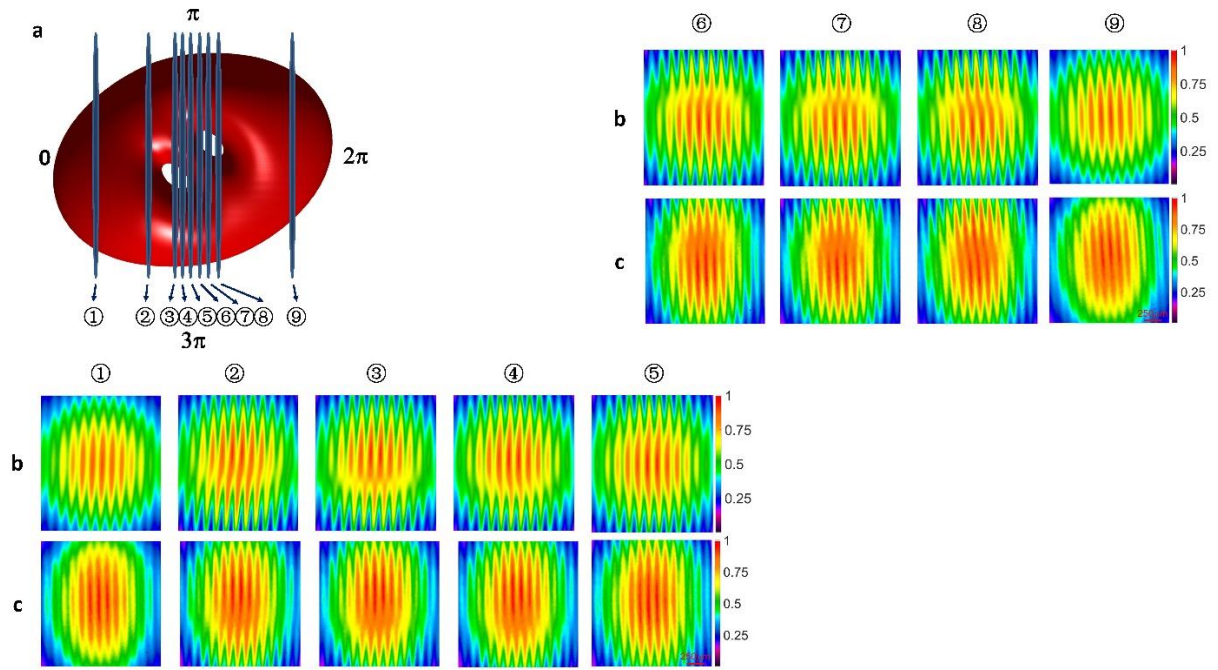

Figure 3. Measurement of the ST vortex for $l = 2$, a) Theoretical ST vortex profile, b) theoretical phase fringe patterns at various temporal locations, c) experimental fringe patterns at various location.

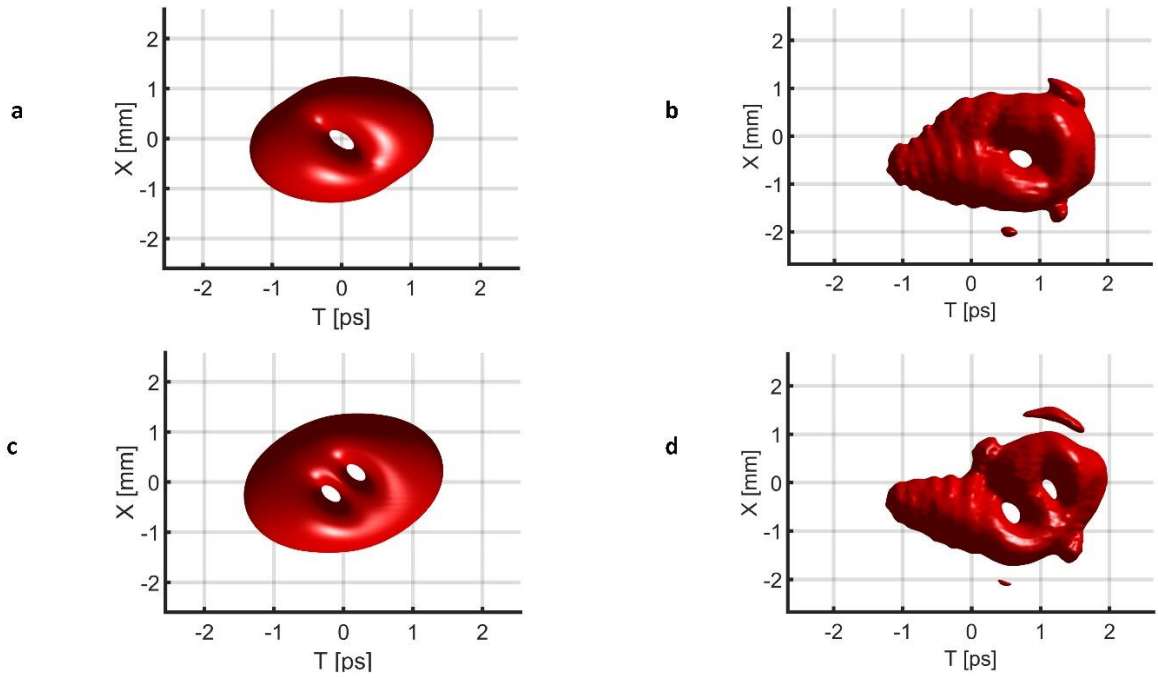

Figure 4. a) Theoretical and b) measured 3D profiles of ST vortex with $l = 1$, c) theoretical and d) measured 3D profiles of ST vortex with $l = 2$.

# Supplementary information of
# Photonic Cyclone: spatiotemporal optical vortex with transverse controllable orbital angular momentum

## I. Supplementary Theory

Starting with an electric field of $g_R(r)$ in the spatial frequency $(k_x)$ – frequency $(\omega)$ domain with $r = \sqrt{k_x^2 + \omega^2}$, applying a phase of $e^{il\theta}$, where $\theta = \tan^{-1}\left(\frac{\omega}{k_x}\right)$ gives the electric field of $f(r,\theta) = g_R(r)e^{il\theta}$. Therefore, $e^{il\theta}$ represent a spiral phase in the $k_x - \omega$ plane with $l$ is an integer referred to as a topological charge. Two-dimensional (2D) Fourier transform of the field is as follows.

$$G(\rho, \phi) = F.T.\{g_R(r)e^{il\theta}\} = \int_0^\infty r\, dr \int_0^{2\pi} d\theta\, g_R(r) e^{il\theta} \exp(i2\pi\rho r \cos(\theta - \phi))$$

where $\rho = \sqrt{x^2 + t^2}$ and $\phi = \tan^{-1}\left(\frac{t}{x}\right)$. The analytical solution of the 2D Fourier transform is as follows [1].

$$G(\rho, \phi) = F.T.\{g_R(r)e^{il\theta}\} = 2\pi\, (-i)^l e^{il\phi} H_l\{g_R(r)\}. \qquad (1)$$

$H_l\{g_R(r)\} = \int_0^\infty r\, g_R(r) J_l(2\pi\rho r) dr$ is the $l$th order Hankel transform where $J_l$ is the $l$th order Bessel function of the first kind. Eqn. 1 indicates that the field with a spiral phase of $e^{il\theta}$ in the $k_x - \omega$ domain is transformed into a field with a spiral phase $e^{il\phi}$ in the $x - t$ domain. The topological charge and therefore the orbital angular momentum (OAM) is preserved under the Fourier transform. Consequently, the wave packet will contain a phase singularity in the $x - t$ plane as a spatiotemporal (ST) vortex

The Hankel transform represents the electric field profile in the $x - t$ plane. For a Gaussian electric field profile of $g_R(r) = \exp(-r^2)$, some Hankel transforms can be obtained analytically. For example, for $l = 1$,

$$H_1\{\exp(-r^2)\} = (\pi^{3/2}\rho/4) \exp\left(-\frac{(2\pi\rho)^2}{8}\right)\left\{I_0\left(\frac{(2\pi\rho)^2}{8}\right) - I_1\left(\frac{(2\pi\rho)^2}{8}\right)\right\},$$

where $I_0$ and $I_1$ are modified Bessel functions of the first kind.

For $l = 2$,

$$H_2\{exp(-r^2)\} = \frac{1}{2(2\pi\rho)^2} exp\left(-\frac{(2\pi\rho)^2}{4}\right)\left\{4 exp\left(\frac{(2\pi\rho)^2}{4}\right) - (2\pi\rho)^2 - 4\right\}.$$

Fig. S1 shows iso-intensity profiles of Eqn. 1 with a Gaussian profile in $y$ – direction for $l = 1$ and $l = 2$. The iso-intensity profile clearly shows a zero intensity region, where the phase singularity exists, near the center of the wave packet. As $l$ increases, the optical energy is spread further in the $x - t$ domain.

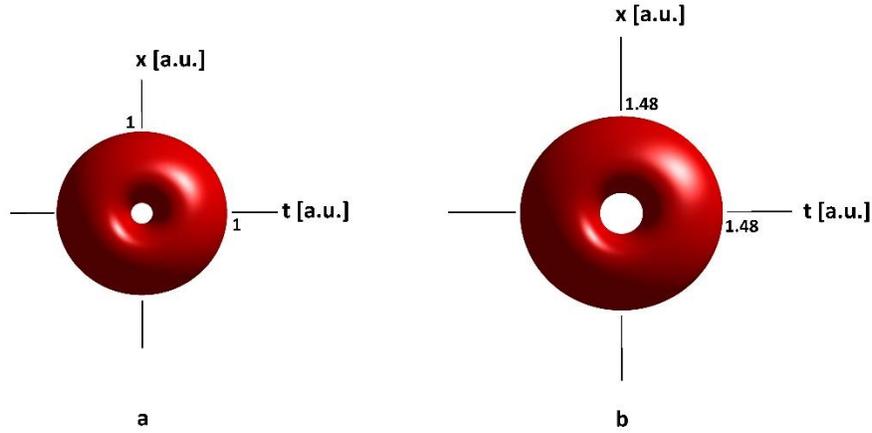

Figure S1. a) Iso-intensity profile of Eqn. 1 with $l = 1$, b) Iso-intensity profile of Eqn.1 with $l = 2$.

## II.  Supplementary method

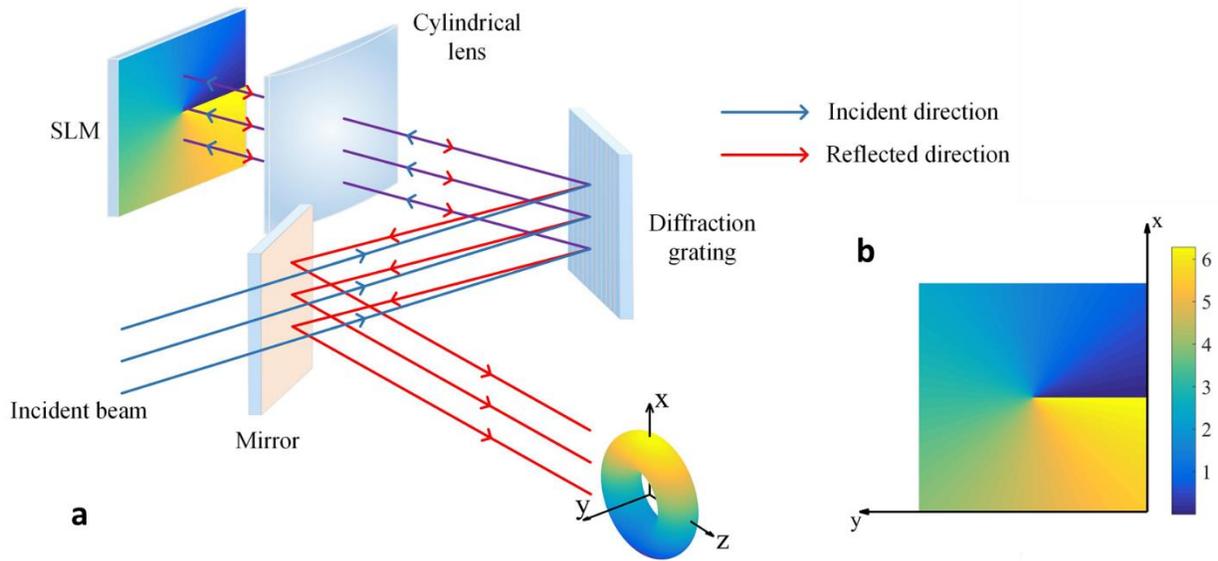

Figure S2. a) Experimental setup to generate a ST vortex, b) a typical phase on the SLM to generate ST vortex. The given phase pattern is for $l = 1$.

Fig. S2 shows the experimental setup schematic, a pulse shaper with a 2D spatial light modulator (SLM) to generate a ST vortex. A diffraction grating and a cylindrical lens spatially spread frequencies of an ultrashort pulse. This process can be understood as a temporal Fourier transform. Hence, one can consider the SLM plane as the $k_x - \omega$ plane. First, a spiral phase is applied to the SLM to form $g_R(r)e^{il\theta}$. By an inverse propagation through the grating-lens, which is an inverse temporal Fourier transform, a ST vortex is generated. Fig. S2b shows a typical phase map ($l = 1$) on the 2D SLM to generate a ST vortex. Even though there is no spatial Fourier transform process in the setup, the free space propagation in the pulse shaper is sufficient to form ST vortices completely.

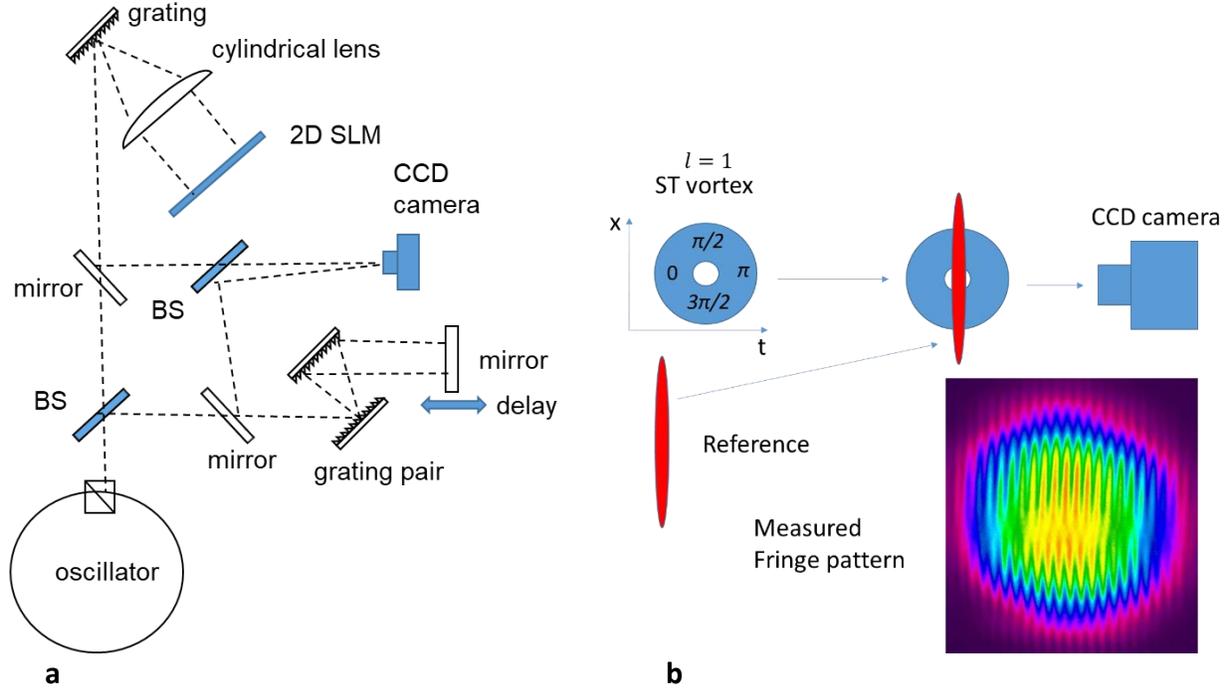

Figure S3. a) Experimental set up to generate and measure ST vortices; BS: beam splitter, b) Schematic of the phase measurement method. Italic numbers represent relative phases for the $l = 1$ ST vortex. Italic numbers represent relative phases at various locations. Note the phase increases in the clockwise direction in $t - x$ plane

Fig. S3 shows the schematics of experimental and measurement setup. Mode-locked pulses from a fiber oscillator are positively chirped (~3ps). While a chirped ST vortex is formed by the spiral phase on the pulse shaper 2D SLM, a short reference pulse (~90 fs) is prepared by compressing the pulse with a grating pair. The ST vortex and the short reference pulse are overlapped with a delay to form a spatial fringe pattern on the CCD camera as shown in Fig. S3b. Spatial fringe images at various delay are collected to reconstruct the phase profile of the ST vortex.

As $e^{il\theta}$ phase is applied in the $x - y$ plane SLM, the phase increasing direction is same in the $x - t$ domain but the direction is reversed in the $t - x$ ($t$ as the horizontal axis while $x$ as the vertical axis) plane. As s result, $l > 0$ spiral phase on the SLM will generate the ST vortex with phase increasing in the clockwise direction in the $t - x$ plane (Fig. 3Sb). We will define the clockwise direction increasing phase in $t - x$ plane as a positive topological charge ($l > 0$) for a ST vortex as shown in Fig. 3Sb.

### III. Supplementary discussion

The fundamental ($l = 1$) ST vortex is stable in propagation. In contrast, higher-order ST vortices experience distortions in propagation. Fig. 4S illustrates the propagation effect in higher-order ST vortices. For example, as a $l = 2$ ST vortex propagates in free space, it experiences the diffraction spatially ($x$-direction in Fig. 4Sa) while there is no dispersion effect in time. As a consequence, up and down energy diffract toward each other and eventually merged constructively since the up and down phase difference is $2\pi$ (Fig. 4Sb). This can be understood as the unbalance between the diffraction and dispersion effect. As a result, two $l = 1$ vortices are formed eventually. In fact, due to the phase evolution in free space propagation, the location of vortices are rotated with respect to the center where the wave packet is merged. The rotational direction depends on the sign of the topological charge.

This is directly analogous to the instability of higher-order optical vortex beams under perturbations [2]. For the ST vortex case, a similar process has been discussed theoretically as a 'diffraction in time' [3].

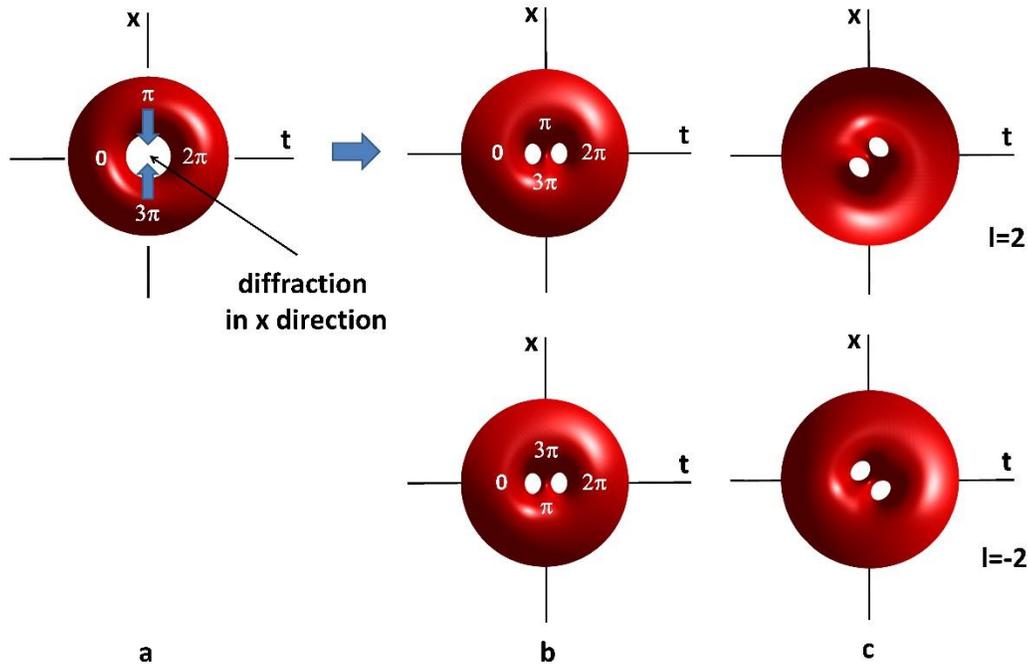

Figure S4. Propagation of $l = \pm 2$ ST vortices. a) Initial $l = 2$ ST vortex, b) center merged $l = \pm 2$ ST vortices. Graphics are presented only to illustrate the diffraction effect. Such horizontally lined up vortices do not occur since the merge and the rotation occur simultaneously. c) ST vortex after some free space propagation (center is merged and the vortex locations are rotated). Italic letters represent the relative phase at various locations.

## IV. Supplementary data

Video S1. Experimental phase pattern movie as the reference pulse is scanned on the ST vortex with $l = 1$. The first two slides are beam profiles of the ST vortex and reference beam respectively. The scanning step size is ~33 fs. Reference pulse duration is ~90 fs.

Video S2. Theoretical phase pattern movie as the reference pulse is scanned on the ST vortex with $l = 1$.

Video S3. Experimental phase pattern movie as the reference pulse is scanned on the ST vortex with $l = -1$. The scanning step size is ~33 fs.

Video S4. Experimental phase pattern movie as the reference pulse is scanned on the ST vortex with $l = 2$. The scanning step size is ~33 fs.

Video S5. Theoretical phase pattern movie as the reference pulse is scanned on the ST vortex with $l = 2$.

Video S6. Experimental 3D iso-intensity movie of the ST vortex with $l = 1$.

Video S7. Experimental 3D iso-intensity movie of the ST vortex with $l = 2$.

## V. Supplementary references